\documentclass[a4paper, 11pt]{article}
\usepackage{graphicx}				
				
\usepackage{cite}
\usepackage{amsmath}
\usepackage{amsfonts}
\usepackage{dsfont}
\usepackage{mathrsfs}
\usepackage{amssymb}
\usepackage{bbm}
\usepackage{epsfig}
\usepackage[english]{babel}
\usepackage[all]{xy}
\usepackage{color}

\newtheorem{theorem}{Theorem}
\newtheorem{definition}{Definition}

\newtheorem{lemma}{Lemma}
\newtheorem{remark}{Remark}

\begin{document}

\title{\bf Kalman Filtering over Fading Channels:\\ Zero--One Laws and Almost Sure Stabilities\thanks{This work was supported in part by the Knut and Alice Wallenberg
Foundation, the Swedish Research Council, and the NNSF of China under
Grant No. 61120106011. Some preliminary results of  the current manuscript  were briefly  presented  at the 53rd IEEE Conference on Decision and Control in LA, USA, 2014.}
}
\author{Junfeng Wu\thanks{J. Wu is  with the College of Control Science and Engineering, Zhejiang Universiy, Hangzhou 310027, China. Email: jfwu@zju.edu.cn}, Guodong Shi\thanks{G. Shi is with the Research School of Engineering, The Australian National University, ACT 0200, Canberra, Australia.    Email:  guodong.shi@anu.edu.au.}, Brian D. O. Anderson\thanks{B. D. O. Anderson is with the Research School of Engineering, Australian National University, and Data61-CSIRO,
  Canberra, ACT 0200, Australia.   Email:  brian.anderson@anu.edu.au.}, and
Karl Henrik Johansson\thanks{K. H. Johansson is with the ACCESS Linnaeus Centre, School of Electrical Engineering, Royal Institute of Technology, Stockholm 10044, Sweden. Email: kallej@kth.se}
 }
\date{}
\maketitle
\begin{abstract}
In this paper, we investigate
probabilistic stability of Kalman filtering over
fading channels modeled by $\ast$-mixing random processes,  where channel fading is allowed to generate non--stationary packet dropouts with temporal and/or spatial correlations.
Upper/lower almost sure (a.s.)
stabilities and absolutely upper/lower a.s. stabilities are defined for characterizing the sample--path behaviors of the Kalman filtering. We prove that both upper and lower a.s. stabilities follow a zero--one law, i.e.,
these stabilities must happen with a probability either zero or one,
and when the filtering system is one--step observable, the absolutely upper and lower a.s. stabilities can also be interpreted using  a zero--one law. We establish general  stability conditions for (absolute) upper and lower a.s. stabilities. In particular, with one--step observability,  we show the equivalence between absolutely a.s. stabilities and a.s. ones, and  necessary and sufficient conditions in terms of packet arrival rate are derived; for the so--called non--degenerate systems, we also manage to give a necessary and sufficient  condition for upper a.s. stability.
\end{abstract}

\section{Introduction}

\subsection{Background and Motivation}
The last decade has witnessed an increasing attention on wireless sensor networks (WSNs) from the control, communication and networking communities, thanks to a rapid development of micro--electronics, wireless communication, and information and networking technologies.
WSNs have applications in a wide range of areas such as health care, intelligent buildings, smart transportation
and power grid,
just to name a few, due to considerable advantages,
including reducing operational cost, allowing distributed
sensing and information sharing among different nodes, etc.
New challenges have also been introduced at the expense of the aforementioned advantages, where control and estimation systems have to be  sustainable in the presence of communication links. This has attracted significant  attention to the  study of information theory for network systems  \cite{Franceschetti2014}, and one fundamental aspect lies in that channel fading~\cite{rappaport1996wireless} leads to constructive or destructive interference of
telecommunication signals, and at times severe drops in the channel signal--to--noise ratio may cause temporary communication outage for the underlying control or estimation systems.

The Kalman filter~\cite{lim2000kalman,chaudhari2010energy} plays a fundamental role in networked  state estimation systems, where a basic  theme  is the stability of Kalman filtering over a communication channel between the plant and the estimator which generates random packet dropouts~\cite{HespanhaNaghshtabriziXu2007}.
There were mainly two stability categories in the literature focusing  on the mean--square, or the  probability distribution,  evolution of the  error covariance along sample--paths of the Kalman filtering, respectively. The majority of the research works assumes the  channel admits  identically and independently distributed (i.i.d.) or Markovian packet drops.
In~\cite{Sinopoli2004}, Sinopoli et al.
modeled the packet losses as an i.i.d. Bernoulli process, and  proved that there exists a critical arrival rate for
the packet arrival rate, below which, the expected prediction error covariance  is unbounded.
Further improvements of this  result  were developed in~\cite{xiangheng,Plarre09tac,shi-tac10}.
The mean--square stability, and stability defined at random packet recovery/reception times,  of Kalman filtering subject to Markovian packet losses generated by a Gilbert--Elliott channel were studied in \cite{huang-dey-stability-kf,xie2008stability,xiao2009kalman,you2011mean,
yilin12criticalvalue,quevedo2013tac-state}.
Efforts have also  been made from a probabilistic point of view.
Weak convergence of Kalman filtering with intermittent observations,  which amounts to having the  error covariance matrix  converge to a limit distribution, were investigated in  \cite{kar2012kalman,censi2011kalman,xie2012stochastic}  for
i.i.d., semi--Markov, and Markovian packet drop models, respectively.
The weak convergence of distributed Kalman filtering was studied in~\cite{Li&Kar14TIT}.

In this paper, we aim to   characterize the  asymptotic  behaviors of the sample paths of Kalman filtering over  fading channels. Instead of only focusing  on certain average property (mean--square, or distribution) of the sample paths, we go beyond most of the stability notions considered in the literature.  It turns out that the majority of the packet drop models can be put  under a unified model from the mixing theory.

\subsection{Model and Contribution}
We assume that the
data packets are regarded as successfully received when received error--free;
and are regarded as completely  lost otherwise.
Although real digital communication introduces a bunch of other challenges, such as quantization and data rate, bit errors, and random delays~\cite{Franceschetti2014}, we are exclusively devoted to studying the impact of
packet dropouts on the estimation performance and
therefore those other effects will be ignored.
To address non--stationarity of the propagation
environment with spatial and temporal correlations between
channel parameters~\cite{zhangJSAC99,
Stojanovic08WONS,dib2009vehicle},
we introduce a packet drop process that is $\ast$-mixing~\cite{blum1963strong}.
The mixing theory
provides a tool of investigating random processes which are  approximately independent in the sense that the dependence dies away as the distance of any two random variables in the process grows large. The $\ast$-mixing model includes but also generalizes i.i.d. and
Markov--type models in the literature.

 We  consider the probabilistic stabilities  of Kalman filtering over such general fading channels. We devise the definitions of upper/lower a.s. stabilities
and absolutely upper/lower a.s. stabilities. The difference and connection between mean--square stability
and (absolutely) a.s. stabilities are  also discussed.
Consistent
with a.s. convergence, the definitions of (absolutely) a.s. stabilities serve  as a supplement of the stability study on
Kalman filtering from the perspective of
probabilistic behaviors. We establish the following results:
\begin{itemize}
\item We prove that the upper and lower a.s. stabilities follow a zero--one
law, indicating that an event
must happen with probability either zero or one.
When the considered filtering
system is one--step observable, the absolutely upper and lower a.s. stabilities
can also be interpreted by the zero--one law.

\item We further present stability conditions for the (absolutely) upper and
lower stabilities. We first give sufficient/necessary conditions
 for general linear time--invariant (LTI) systems. One--step observable systems yield tighter
 results with
necessary and sufficient conditions in terms of the packet arrival rate
derived for upper and lower a.s.
stabilities. It is also shown for one--step observable systems that
a.s. stability is equivalent to absolutely a.s. one.
Finally, for the so--called non--degenerate systems,
we manage to give a necessary and sufficient
upper a.s. stability condition.
\end{itemize}

All the above results are established under
$\ast$-mixing fading channels, and to the best of our knowledge, this is the first
time the concept of mixing has been
introduced to the  modelling of random packet losses.  An embryo of part of this work (some stability conditions) was presented in~\cite{Wu14cdc} for independent channels.

\subsection{Paper Organization}
The remainder of the paper is organized as follows. Section~\ref{section:problem-setup}
provides the problem setup, defines the (absolutely) upper/lower
a.s. stabilities, and introduces the $\ast$-mixing random process considered in~\cite{blum1963strong}.
The difference between various stabilities are also discussed in
Section~\ref{section:problem-setup}. In
Section~\ref{section:Converge-or-Diverge}, two stability zero--one laws are
derived. Various stability conditions are studied
in Section~\ref{section:stability-conditions}.
Some concluding remarks are
given in the end.

\textit{Notations}:
$\mathbb{N}$ is
the set of positive integers.
For a real number $x$, $\lfloor{x}\rfloor$ and
$\lceil{x}\rceil$ denote the largest integer not greater than $x$ and
 the smallest integer not less than $x$ respectively. The set of $n$ by $n$ symmetric positive semi--definite (positive definite) matrices over
the complex field is denoted as $\mathbb{S}_{+}^{n}$ ($\mathbb{S}_{++}^{n}$). For a matrix $X$,
$X'$ denotes the transpose of $X$, and $X^*$ represents the conjugate transpose of
$X$. Moreover, $\lambda_i(X),~i=1,\ldots,n,$
represents the $i$th largest eigenvalue of $X$ in terms of magnitude.
and  $\|X\|_2$ means the spectral norm of $X$.
The indicator function of a subset $\mathcal{A}\subset\Omega$ is a function
${1}_\mathcal{A}:\Omega \rightarrow \{ 0,1 \}$, where
${1}_\mathcal{A}(\omega)=1$ if $\omega\in \mathcal{A}$, otherwise
${1}_\mathcal{A}(\omega)=0$. $\sigma(\cdot)$ denotes
the $\sigma$--algebra generated by
random variables. {For an event $\mathcal A$ in some probability space,
``$\mathcal A~i.o.$ '' means $\mathcal A$ happens infinitely often.}


\section{Kalman Filtering over
Fading Channels}\label{section:problem-setup}

In this section, we introduce the Kalman filtering model and define the problem of interest.
\subsection{Kalman Filtering with Packet
Dropouts}
Consider an LTI system:
\begin{eqnarray}
x_{k+1} & = & Ax_k + w_k, \label{eqn:process-dynamics} \\
y_k & = & Cx_k + v_k, \label{eqn:measurement-equation}
\end{eqnarray} where $x_k \in \mathbb{R}^{n}$ is the process state vector, $y_k \in \mathbb{R}^{m}$ is the observation vector, $w_{k} \in \mathbb{R}^{n} $ and
$v_k \in \mathbb{R}^{m}$ are zero--mean Gaussian random vectors
with $\mathbb{E}[w_{k}w_{j}{'}] = \delta_{kj}Q~(Q\geq 0)$,
$\mathbb{E}[v_{k}v_{j}{'}] = \delta_{kj}R~(R > 0)$, and
$\mathbb{E}[w_{k}v_{j}{'}] = 0 \; \forall j,k$. The $\delta_{kj}$ is the Kronecker
delta function with $\delta_{kj} = 1$ if $k=j$ and $\delta_{kj}=0$ otherwise.
The initial state $x_0$ is a zero--mean Gaussian random vector that is uncorrelated
with $w_k$ and $v_k$ and has covariance $P_0\geq 0$.
We assume that the pair $(C,A)$ is observable and $(A,Q^{1/2})$ controllable.
We introduce the standard definition of observability index of the pair $(C,A)$.
\begin{definition}\label{def:observability-index}
For the observable pair $(C,A)$, the \textit{observability index} $\mathrm{I_o}\in\mathbb{N}$ is defined as
the smallest integer such that $[C',A'C',\ldots,(A^{\mathrm{I_o}-1})'C']'$ has
 full column rank.
\end{definition}
It is evident that $\mathrm{I_o}\leq n$.

Purely stable LTI systems do not interest us as their
estimation error covariance matrix automatically decays. In that case, it becomes trivial to discuss probabilistic stability  issues.
For unstable LTI systems, it can be seen that, by applying a similarity transformation, the unstable and stable modes can be decoupled. An open--loop prediction for the stable mode always has a bounded estimation error covariance, therefore, this mode does not play any key role in the problem considered here. Without loss of generality, we assume that
\begin{enumerate}
\item[(A1)]\label{asmpt:assumpt-A1}
\emph{All of the eigenvalues of $A$ have magnitudes no less than 1.}
\end{enumerate}

\setlength{\unitlength}{1.5mm}
\begin{figure}[htbp]
\thicklines
\centering
\begin{picture}
(56,14)(0,-7)
\thicklines
\put(-3.5,0){\vector(1,0){3.5}}
\put(0,-2){\framebox(8,4){${\rm Process}$}}
\put(13.5,-2){\framebox(8,4){${\rm Sensor}$}}
\put(43,-2){\framebox(11,4){${\rm Estimator}$}}
\put(8,0){\vector(1,0){5}}
\put(21.5,0){\vector(1,0){4.5}}
\put(26,-4){\dashbox(12,8)}
\put(28.3,1){$\rm Erasure$}
\put(28,-2.5){$\rm Channel$}
\put(38,0){\vector(1,0){5}}
\put(54 ,0){\vector(1,0){3.5}}
\put(55,2.5){$\hat x_{k|k}$}
\put(17 ,6){\vector(0,-1){4}}
\put(-3,2.5){$w_k$}
\put(9,2.5){$x_k$}
\put(22.5,2.5){$y_k$}
\put(18,4){$v_k$}
\put(38.5,2.5){$\gamma_ky_k$}
\end{picture}
\caption{State estimation over an erasure channel.}
\label{fig:block-diagram}
\end{figure}
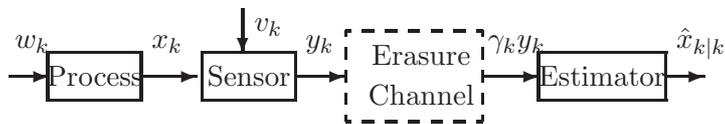

We consider an estimation scheme where
the raw measurements $\{y_k\}_{k\in\mathbb{N}}$ of the sensor
are transmitted to the estimator over
an erasure communication channel over which packets may be
dropped randomly, see Fig.~\ref{fig:block-diagram}. We assume that the packets are regarded as successfully received when received error--free;
and are regarded as completely lost otherwise.
Denote by $\gamma_k\in\{0,1\}$ the arrival of $y_k$ at time $k$:
If $\gamma_k=1$, it indicates that $y_k$ successfully arrives at the estimator; otherwise
$\gamma_k=0$.
We assume that the sequence $\{\gamma_k\}_{k\in {\mathbb{N}}}$
is independent of how the system evolves, and that the estimator knows whether the packet is arrived or not at each time.
Define $\mathcal{F}_k$ as the filtration generated by all the measurements received by
the estimator up to time $k$, i.e.,
$\mathcal{F}_{k}\triangleq \sigma\hspace{-0.5mm}\left(\gamma_ty_t,\gamma_t;1\leq t\leq k\right)$, and define $\mathcal{F}=\sigma\left(\cup_{k=1}^\infty \mathcal{F}_k\right)$. We use a triple $(\Omega,\mathcal{F},\mathbb{P})$ to denote the common probability space for all random variables.
The estimator computes $\hat{x}_{k|k}$, the minimum mean--squared error estimate,
and $\hat x_{k+1|k}$, the one--step prediction, according to
$\hat{x}_{k|k} =\mathbb{E}[x_k|\mathcal{F}_k]$ and
$\hat{x}_{k+1|k} =\mathbb{E}[x_{k+1}|\mathcal{F}_k]$, where
$\mathbb{E}$ denotes the expectation induced by $\mathbb{P}$.
 Let $P_{k|k}$ and $P_{k+1|k}$ be the corresponding estimation and prediction error covariance matrices, receptively, i.e.,
$P_{k|k}=\mathbb{E}[(x_k-\hat x_{k|k}) (\cdot)'|\mathcal{F}_k]$
and $P_{k+1|k}=\mathbb{E}[(x_{k+1}-\hat x_{k+1|k}) (\cdot)'|\mathcal{F}_k]$,
which are computed recursively via a modified Kalman filter~\cite{Sinopoli2004}:
\begin{align*}
  K_{k} & =  P_{k|k-1}C'(CP_{k|k-1}C' + R)^{-1},  \\
  \hat{x}_{k|k} & =  \hat{x}_{k|k-1} + \gamma_k K_{k}(y_{k} - CA\hat{x}_{k|k-1}),  \\
  P_{k|k} & = (I -\gamma_k K_{k}C)P_{k|k-1}, \\
  \hat{x}_{k+1|k}& =  A\hat{x}_{k|k},  \\
  P_{k+1|k} & =  AP_{k|k}A' + Q.
\end{align*}
In particularly, $P_{k+1|k}$ evolves in the following way
\begin{align}
P_{k+1|k}&=AP_{k|k-1}A'+Q\nonumber\\
&-\gamma_kAP_{k|k-1}C'(CP_{k|k-1}C'+R)^{-1}
CP_{k|k-1}A'.
\end{align}

It can be seen that $P_{k+1|k}$ now becomes a function of the random variables $\{\gamma_t\}_{1\leq t\leq k}$. In what follows, we are devoted to
characterizing the impacts of $\{\gamma_k\}_{k\in {\mathbb{N}}}$ on $P_{k+1|k}$.
To simplify discussion in the sequel, let us use a simpler notation
$P_{k+1}\triangleq P_{k+1|k}$, and
introduce the functions $h$, $g$, $h^k$ and $g^k$: $\mathbb{S}^n_+ \to \mathbb{S}^n_+$ as
follows:
\begin{align}
h(X)&\triangleq AXA^{'}+Q,\label{def:h-function}\\
g(X)&\triangleq AXA'+Q-AXC^{'}
{(CXC^{'}+R)^{-1}CXA'},\label{def:g-function}
\end{align}\vspace{2mm}
$h^k(X)\triangleq \underbrace{h\circ h \circ \cdots \circ h}_{k \hbox{~times}}(X)$ and
$g^k(X)\triangleq \underbrace{g\circ g \circ \cdots \circ g}_{k \hbox{~times}}(X)$,
where $\circ$ denotes the function composition.

\subsection{$\ast$-mixing Fading Channels}\label{sec:mixing-fading-channels}
Wireless channels are mainly affected by path loss, small--scale fading and
shadow fading. In a wireless connected vehicle--to--vehicle network~\cite{dib2009vehicle}, for example,
for the sake of moving vehicles, small--scale fading happens
in an unpredictable way. Moreover, shadowing fading, caused by
objects obstructing, leads to temporal and spatial correlations
between communications links. The aforementioned factors are no longer negligible.
To model packet dropouts subject to spatially and/or
temporally correlated and non--stationary fading channels,
 on one hand, we need to take the non--stationarity
of propagation environment and correlations between channel parameters into account;
on the other hand, we have to retain indispensable assumptions,
making it possible to build up instructive theories upon it.
We model the packet dropouts as a $\ast$-mixing stochastic process, where the concept of mixing, originating from
physics, is an attempt to interpret the thermodynamic behavior of mixtures.

Before proceeding, we introduce the definition of $*$-mixing, which is
taken from~\cite{blum1963strong}.
\begin{definition}\label{def:mixing}
The sequence of random variables $\{\xi_k\}_{k\in\mathbb{N}}$ on a probability space $(\mathscr{S},\mathcal{S},\mu)$ is said to be $*$-mixing if there exists
a positive integer $\mathrm{N}$ and a real--valued function $f$ defined for $n\geq \mathrm{N}$,
{where
$n\in\mathbb N$,} such that
\begin{enumerate}
\item[(i)] $f$ is a non--increasing function with $\lim_{n\to\infty}f(n)=0$;
\item[(ii)] There holds $\big|\mu(\mathcal{A}\cap\mathcal{B})-
\mu(\mathcal{A})\mu(\mathcal{B})\big|\leq f(n)\mu(A)\mu(B)$ for all $n\geq \mathrm{N}$, $\mathcal{A}\in
\sigma(\xi_1,\ldots,\xi_k),\mathcal{B}\in\sigma(\xi_{k+n},\xi_{k+n+1},
\ldots)$, and $k\in\mathbb{N}$.

\end{enumerate}
\end{definition}

In the sequel, we assume that
\begin{enumerate}
\item[(A2)]\label{asmpt:assumpt-A2}
\emph{The random process $\{\gamma_k\}_{k\in\mathbb{N}}$ is $*$-mixing.}
\end{enumerate}

To the best of our knowledge, this is the first
time mixing has been introduced when modelling random
packet dropouts. One coarse way to explain the above mathematical definition is that $*$-mixing
implies that the occurrence of any two groups of possible
states can be considered approximately independent
as long as the two groups are a sufficient amount of time apart from each other, where dependence is ``quantified''
by the mixing coefficient $f(n)$.
It is a universal understanding that in the physical world historical states in remote past impact less and less on the evolution of future states, provided
that the hypothesis of $*$-mixing stands. Note that
the idea of mixing has been used in the ``theoretical channel model''
(the theoretical channel refers to a mapping from the input source to
the output source) in the literature~\cite{Alajaji94TIT,Sethuraman05TIT,gray2011entropy}.

Remarkably enough the mixing model admits most of the well--studied models
reported in the literature, e.g., i.i.d.~\cite{Sinopoli2004,Plarre09tac,kar2012kalman}, Markov~\cite{Shi10tac,yilin12criticalvalue,you2011mean,xie2012stochastic},
semi--Markov\cite{censi2011kalman,quevedo2013tac-state}, Markovian jump~\cite{xiao2009kalman}, finite--state Markov~\cite{zhang1999finite},
as its special cases~\cite{blum1963strong}.

\subsection{Problems of Interest}
In this paper, we are interested in  the sample--path behaviors of Kalman filtering with $\ast$-mixing packet losses. Since
$\mathrm{Tr}(P_k)$ represents the sum of squared error variance of
the estimate for each element of $x_k$, we may without loss of generality use
$\mathrm{Tr}(P_k)$ as a performance metric.
Noting that $\limsup_{k\to \infty } \mathrm{Tr}(P_k)$ and $\liminf_{k\to \infty } \mathrm{Tr}(P_k)$ are well--defined random variables taking values from $\mathbb{R} \cup \{+\infty\}$, we  introduce the following stability notions for the considered Kalman filter.
\begin{definition}\label{def:as-statbility}
The considered  Kalman filter is termed
\begin{itemize}
\item[(i)] upper a.s. stable if $\mathbb{P}\big(\limsup_{k\to \infty } \mathrm{Tr}(P_k) < \infty\big)=1$, and lower a.s. stable if $\mathbb{P}\big(\liminf_{k\to \infty } \mathrm{Tr}(P_k) < \infty\big)=1$;

\item[(ii)] absolutely upper a.s.  stable if there exists a constant $\mathrm{C}>0$ such that $
\mathbb{P}\big(\limsup_{k\to \infty } \mathrm{Tr}(P_k) < \mathrm{C}\big)=1
$,
 and absolutely lower a.s. stable if there exists a constant $\mathrm{C}>0$ such that $
\mathbb{P}\big(\liminf_{k\to \infty } \mathrm{Tr}(P_k) < \mathrm{C}\big)=1.
$
\end{itemize}
\end{definition}

These stability notions  focus on
the asymptotic behavior of the estimation  system along every sample path across
the sample space,
enabling us to investigate a Kalman filtering system from  a probabilistic perspective. Note that, in general, absolutely a.s. stability is a stronger notion than the a.s. one. For convenience, we also call the considered  Kalman filter  {\it upper a.s. unstable} if $
\mathbb{P}\big(\limsup_{k\to \infty } \mathrm{Tr}(P_k) < \infty\big)=0$,
and
{\it lower a.s.  unstable} if $
\mathbb{P}\big(\liminf_{k\to \infty } \mathrm{Tr}(P_k) < \infty\big)=0$. Additionally,  the Kalman filter is said to be almost surely convergent if
\begin{align}\label{guodong2}
\mathbb{P}\Big(\lim_{k\to \infty } P_k \hbox{~exists,~and~is~finite}\Big)=1.
\end{align}


\subsection{Discussions}\label{section:diff-statbility-discussion}
In the literature, a widely investigated stability notion of  Kalman filtering systems
with packet losses is mean--square stability, i.e., the Kalman filtering is mean--square stable if  $\sup_{k\in\mathbb{N}} \mathbb{E} \| P_{k}\|<\infty$. In general, there are no implications between a.s./absolutely a.s.
 stabilities  and mean--square stability  for the Karlman filter. This relation is analogous to the relation between a.s. convergence and $L_2$--convergence for a sequence of random variables \cite{durrett2010probability}, because  a.s./absolutely a.s.
 stabilities are defined on the basis of a.s. convergence and
 mean--square stability is defined on the basis of $L_2$--convergence.

 Another important concept of Kalman filtering systems  is the weak--convergence, which requires $P_k$ to converge to a limit in distribution \cite{kar2012kalman,censi2011kalman,xie2012stochastic}. Then by standard chain of implications of the notions of probabilistic  convergence, we know that both the mean--square convergence (i.e., $\lim_{k\to \infty }\mathbb{E} \| P_{k} -P_\ast \|=0$ for some $P_\ast$) and the a.s. convergence (\ref{guodong2}) imply  weak convergence.

\section{The Zero--One Laws}\label{section:Converge-or-Diverge}

%
%
%

 A tail event of a random process is an event whose
 occurrence is independent of each
 finite subset of random variables.  In this section, we present that
 the a.s. stabilities follow a
 zero--one law, which is shown with the aid of the definition of tail events and
the zero--one law for a $*$-mixing sequence.
%
\begin{theorem}\label{thm:zero-one-law} Let Assumptions (A1)--(A2) hold.
Both upper and lower a.s. stabilities follow a zero--one law, i.e.,
\begin{itemize}
\item[(i)] Either $\mathbb{P}\big(\limsup_{k\to \infty } \mathrm{Tr}(P_k) < \infty\big)=1$ or $\mathbb{P}\big(\limsup_{k\to \infty } \mathrm{Tr}(P_k) < \infty\big)=0$;
\item[(ii)] Either $\mathbb{P}\big(\liminf_{k\to \infty } \mathrm{Tr}(P_k) < \infty\big)=1$ or $\mathbb{P}\big(\liminf_{k\to \infty } \mathrm{Tr}(P_k) < \infty\big)=0$.
\end{itemize}
\end{theorem}


The following theorem further shows  that the zero--one law also applies to
absolutely upper and lower a.s. stabilities when the system is one--step observable, i.e., $\mathrm{I_o}=1$.

\begin{theorem}\label{thm:zero-one-law-2} Let Assumptions (A1)--(A2) hold.
Suppose $\mathrm{I_o}=1$. Then  absolutely upper and lower  a.s. stabilities follow the zero--one law, i.e.,
\begin{itemize}
 \item[(i)] Either there exists a constant $\mathrm{C}>0$ such that $\mathbb{P}\big(\limsup_{k\to \infty } \mathrm{Tr}(P_k) < \mathrm{C}\big)=1$ or \\
     $\mathbb{P}\big(\limsup_{k\to \infty } \mathrm{Tr}(P_k) < \mathrm{C}\big)=0$ holds for any $\mathrm{C}>0$;
     \item[(ii)] Either there exists a constant $\mathrm{C}>0$ such that $\mathbb{P}\big(\liminf_{k\to \infty } \mathrm{Tr}(P_k) < \mathrm{C}\big)=1$ or \\$\mathbb{P}\big(\liminf_{k\to \infty } \mathrm{Tr}(P_k) < \mathrm{C}\big)=0$ holds for any $\mathrm{C}>0$.
         \end{itemize}
\end{theorem}

In the rest of this section, we first gather and establish a few supporting
lemmas, and then provide detailed proofs for Theorems~\ref{thm:zero-one-law}~and~\ref{thm:zero-one-law-2}.

\subsection{Supporting Lemmas}

Denote the unique solution to $g(X)=X$ as $\overline P$. Assuming the observability of $(C,A)$  and controllability of $(A,Q^{1/2})$,
it is well known that $\overline P$ is a positive definite matrix~\cite{lancaster1995algebraic};
and that, for a standard Kalman filter, $\lim_{k\rightarrow \infty}P_{k}=\overline P$~\cite{anderson79}.
For the operators $h$ and $g$, the following lemma holds.
The proof can be found in Lemma A.1 in~\cite{shi-tac10}.
\begin{lemma}\label{lemma:appendix-g-h}
For any matrices $X\geq Y\geq 0$,
\begin{eqnarray}
h(X)&\geq& h(Y),\label{eqn:h-monotone}\\
g(X)&\geq& g(Y),\label{eqn:g-monotone}\\
h(X)&\geq& g(X).\label{eqn:h-geq-g}
\end{eqnarray}
\end{lemma}

The following two  lemmas further establish some useful  properties of operators $g$ and $h$.
\begin{lemma}\label{lemma:appendix-positive-definite}
For any $X\in\mathbb{S}_+^n$, there exists an integer $t\in \mathbb{N}$, independent of $X$, such that $g^t(X)>0$.
\end{lemma}{
{\it Proof.} Choose a constant $\beta\in(0,1)$.
Since $\lim_{k\rightarrow\infty}g^k(0)=\overline{P}$,  there always exists
a sufficiently large integer $\mathrm{N}(\beta)$ such that $g^k(0)\geq \beta\overline{P}>0$ for all $k\geq \mathrm{N}(\beta)$. Then
By Lemma~\ref{lemma:appendix-g-h}, $g^k(X)\geq g^k(0)>0$. Note that $\mathrm{N}(\beta)$ is chosen independent of $X$.
The conclusion follows by letting $t=\mathrm{N}(\beta)$.
\hfill$\square$}

\begin{lemma}\label{lemma:appedix-1}
There exists
 a constant $a>0$ such that
$\mathrm {Tr}\left(h^k(X)\right)\geq a |\lambda_1(A)|^{2k}$ holds
for all $X\in \mathbb{S}_+^n$   and for all $k\in\mathbb N$.
\end{lemma}
{\it Proof.}
By
the controllability of $(A,Q^{1/2})$ assumed, one has $V\triangleq h^n(0)>0$.
Then there always exists a real number $a_0>0$ so that $V\geq a_0 I$. Therefore,
$h^k(0)\geq a_0 A^{k-n}(A')^{k-n}$ holds for all $k\geq n$. Let us denote the Schur's unitary triangularization~\cite{horn2012matrix}
of $A$ as $A=UTU^*$ where $U$ is a unitary matrix and $T=[t_{ij}]$ is an upper triangular with
$t_{ii}=\lambda_i(A),~i=1,\ldots,n.$ Since $A^{k-n}(A')^{k-n}$ is Hermitian and positive semi--definite,
$\lambda_1\left(A^{k-n}(A')^{k-n}\right)$ is real and moreover,
\begin{align*}
\lambda_1\left(A^{k-n}(A')^{k-n}\right)
&=\lambda_1\left(T^{k-n}(T^*)^{k-n}\right)\\
&=
\Bigg\|\left[\begin{array}{ccc}
\lambda_1(A^{k-n}) & * & *\\
0                  &\ddots&*\\
0                  & 0 &\lambda_n(A^{k-n})
\end{array}\right]\Bigg\|_2^2\\
&\geq |\lambda_1(A^{k-n})|^{2}\\
&=
|\lambda_1(A)|^{2(k-n)}.
\end{align*}
Therefore, $\mathrm {Tr}\left(h^k(0)\right)\geq a_n|\lambda_1(A)|^{2k}$ holds
for all $k\geq n$
with $a_n\triangleq a_0|\lambda_1(A)|^{-2n}$. As for $k=1,\ldots,n-1$, we choose a sequence of
positive real numbers, denoted by $\{a_k\}_{1\leq k\leq n-1}$, such that
$\mathrm {Tr}\left(h^k(0)\right)\geq a_k|\lambda_1(A)|^{2k}$.
The conclusion follows by taking $a\triangleq\min\{a_k:k=1,\ldots,n\}>0$.
\hfill$\square$

Since $(C, A)$ is observable, $J\triangleq \left[
(CA^{\mathrm{I_o}-1})',(CA^{\mathrm{I_o}-2})',\ldots,C'\right]'$
has full column rank and $J'J$ is nonsingular.
Denote
\begin{align}\label{eqn:def-M0}
&M_0\triangleq(J'J)^{-1}J'\left(
H\left[\begin{array}{ccc}
Q& \ldots & 0\\
\vdots & \ddots & \vdots\\
0 & \ldots & Q\end{array}
\right]H'+\left[\begin{array}{ccc}
R& \ldots & 0\\
\vdots & \ddots & \vdots\\
0 & \ldots & R\end{array}
\right]
\right)J(J'J)^{-1}
\end{align}
and
$$
H=\left[\begin{array}{cccccc}
C & CA & \ldots &\ldots & CA^{\mathrm{I_o}-2} \\
0 & C & \ddots & &\vdots \\
\vdots & \vdots & \ddots &\ddots &\vdots \\
0 & 0 & \ldots & C & CA \\
0 & 0 & 0 & \ldots & C \\
0 & 0 & 0 & \ldots & 0\\
\end{array}\right].
$$
Next define
\begin{equation}\label{def:M}
M=h^{\mathrm{I_o}}(M_0).
\end{equation}
For $\mathrm{I_o}$ and $M$, we have the following lemma.
\begin{lemma}\label{lemma:suboptima-linear-estimator}
Suppose that by time $k-1$ there are at least $\mathrm{I_o}$ numbers of consecutive measurements
$y_{k-\mathrm{I_o}},\ldots,y_{k-1}$ received by the Kalman filter. Then there holds
$P_k\leq M$.
\end{lemma}
{\it Proof.}
Observe that
\begin{equation*}
\left[\begin{array}{c}
y_{k-1}\\
y_{k-2}\\
\vdots\\
y_{k-\mathrm{I_o}}
\end{array}\right]=
Jx_{k-\mathrm{I_o}}+
H\left[\begin{array}{c}
w_{k-2}\\
w_{k-3}\\
\vdots\\
w_{k-\mathrm{I_o}}
\end{array}\right]+
\left[\begin{array}{c}
v_{k-1}\\
v_{k-2}\\
\vdots\\
v_{k-\mathrm{I_o}}
\end{array}\right].
\end{equation*}
Based on the consecutive measurements
$y_{k-\mathrm{I_o}},\ldots,y_{k-1}$ received by the estimator,
we use the following estimator to generate a
linear prediction of $x_k$:
$$
\bar x_k=A^{\mathrm{I_o}}(J'J)^{-1}J'\left[\begin{array}{c}
y_{k-1}\\
y_{k-2}\\
\vdots\\
y_{k-\mathrm{I_o}}
\end{array}\right].
$$
The associated prediction error covariance $\overline{P}_k\triangleq
\mathbb{E}[(x_k-\bar x_k)(x_k-\bar x_k)']\leq M$.
Since the Kalman filter is known as the linear minimum mean--squared error estimator,
{we have
$P_k\leq \overline{P}\leq {M}$,} which completes the proof.
\hfill$\square$

In the following, we introduce the definition of Riemannian distance
on $\mathbb{S}_{++}^n$.
\begin{definition}
 For any $X,Y\in \mathbb{S}_{++}^n$, the Riemannian distance $\delta$
 between $X$ and $Y$
is defined as
\begin{equation}\label{def:riemannian-distance}
\delta(X,Y)=\left(\sum_{i=1}^n \mathrm{Log}^2\lambda_i\left(XY^{-1}\right)\right)^{1/2}.
\end{equation}
\end{definition}

It has been shown that $\delta$ is a metric on $\mathbb{S}_{++}^{n}$, and that  the metric space $(\mathbb{S}_{++}^n, \delta)$ is complete and separable \cite{Bougerol1993SIAM}.
In $(\mathbb{S}_{++}^n, \delta)$, the
operators $h,~g$ defined in~\eqref{def:h-function} and~\eqref{def:g-function}   are non--expansive and $g^{\mathrm{I_o}}$ is contractive.
\begin{lemma}
\label{lemma:contraction-mapping} (Theorem 1.7, ~\cite{Bougerol1993SIAM})
Suppose that $A$ is invertible. In the metric space $(\mathbb{S}_{++}^n,\delta)$,
\begin{itemize}
\item[(i)] There hold
$\delta(h(X),h(Y))\leq \delta(X,Y)\hbox{~and~}
\delta(g(X),g(Y))\leq \delta(X,Y)$ for any $X,Y\in\mathbb{S}_{++}^n$;
\item[(ii)] There exists a real number $q\in(0,1)$ that only depends on
$A,~C,~Q,~R$ such that there holds
$$
\delta(g^{\mathrm{I_o}}(X),g^{\mathrm{I_o}}(Y))\leq  q\delta(X,Y)$$
for any $X,Y\in\mathbb{S}_{++}^n$.
\end{itemize}
\end{lemma}

It is also easy to establish the following lemma.
\begin{lemma}\label{lemma:comparable-matrix}
In the metric space $(\mathbb{S}_{++}^n,\delta)$,
there holds $2^{-\delta(X,Y)}X\leq Y\leq 2^{\delta(X,Y)}X$ for any $X,Y\in\mathbb{S}_{++}^n$.
\end{lemma}
{\it Proof.}
From the definition of Riemannian distance in~\eqref{def:riemannian-distance}, we have
$$
\mathrm{Log}\lambda_n(XY^{-1})\leq \delta(X,Y)
\hbox{~and~}
\mathrm{Log}\lambda_1(XY^{-1})\geq -\delta(X,Y).
$$
Therefore, $2^{-\delta(X,Y)}I\leq Y^{-1/2}XY^{-1/2}\leq 2^{\delta(X,Y)}I,$
which completes the proof.
\hfill$\square$

Next, we consider a deterministic  sequence $\{z_k\}_{k\in\mathbb{N}}$ with each $z_k$ taking value from $\{0,1\}$. Associated with the sequence $\{z_k\}_{k\in\mathbb{N}}$  we define the (deterministic) recursion:
\begin{align}\label{deterministic}
P_{k+1}=AP_kA'+Q- z_kAP_kC'(CP_k C'+R)^{-1}
CP_kA'.
\end{align}
The following lemma holds.
\begin{lemma}\label{proposition:limsup-zero-one-law} Consider  the deterministic evolution  (\ref{deterministic}).
\begin{itemize}
\item[(i)] If  there exists an initial condition
$P_0=\Sigma$ such that $\limsup\limits_{k\rightarrow \infty}\mathrm{Tr}(P_k)=\infty$, then\\
$\limsup\limits_{k\rightarrow \infty}\mathrm{Tr}(P_k)=\infty$
for all  $P_0\in\mathbb{S}_+^n$;

\item[(ii)] If there exists an initial condition
$P_0=\Sigma$ such that $\limsup\limits_{k\rightarrow \infty}\mathrm{Tr}(P_k)<\infty$, then\\
$\limsup\limits_{k\rightarrow \infty}\mathrm{Tr}(P_k)<\infty$
for all $P_0\in\mathbb{S}_+^n$.
\end{itemize}
\end{lemma}
{\it Proof.}
Consider two Kalman filters that undergo the packet loss process $\{z_k\}_{k\in\mathbb{N}}$: one has initial condition $\Sigma_1$ while the other has initial condition $\Sigma_2$.
Denote the prediction error covariance matrices at time $k$ from
initial points $\Sigma_1$ and $\Sigma_2$, respectively,
by $P_{k}^{\Sigma_1}$ and $P_{k}^{\Sigma_2}$.
From Lemma~\ref{lemma:appendix-positive-definite}, we can always find a sufficiently large integer
$t$ such that $g^t(\Sigma_1),g^t(\Sigma_2)\in\mathbb{S}_{++}^n$.
According to~\eqref{eqn:g-monotone} and~\eqref{eqn:h-geq-g}, we have $P_{t}^{\Sigma_1}, P_{t}^{\Sigma_2}\in\mathbb{S}_{++}^n$. On the other hand, $P_{t}^{\Sigma_1}\leq h^t(\Sigma_1)$ and $P_{t}^{\Sigma_2}\leq h^t(\Sigma_2)$ by~\eqref{eqn:h-geq-g}.
Therefore, there always exists a constant $d\geq 0$ such that
$\delta(P_{t}^{\Sigma_1}, P_{t}^{\Sigma_2})\leq d$. The fact from Lemma~\ref{lemma:contraction-mapping} that the operators
$h$ and $g$ are non--expansive in $(\mathbb{S}_{++}^n,\delta)$ provided that $A$ is invertible leads to
$$\delta(P_{k}^{\Sigma_1}, P_{k}^{\Sigma_2})\leq d$$
~for~all~$k\geq t$.
By Lemma~\ref{lemma:comparable-matrix},
\begin{equation}\label{eqn:beta-relation}
P_{k}^{\Sigma_1} \geq \beta P_{k}^{\Sigma_2}
\end{equation}
holds for~all~$k\geq t$, where $\beta\triangleq 2^{-d}.$

The unboundness of $\mathrm{Tr}(P_{k}^{\Sigma})$ means that, for
any positive number $\mathrm{C}$, there always exists
a sufficiently large integer $\mathrm{N}\geq t$ such that
$\mathrm{Tr}(P_{\mathrm{N}}^{\Sigma})>\mathrm{C}$.
When taking $\Sigma_1=0$ and $\Sigma_2=\Sigma$
in~\eqref{eqn:beta-relation}, we have
$\mathrm{Tr}(P_{\mathrm{N}}^0)\geq \beta\mathrm{C}$. By~\eqref{eqn:h-monotone} and
\eqref{eqn:g-monotone} again,
 $\mathrm{Tr}(P_{\mathrm{N}}^{P_0})\geq \mathrm{Tr}(P_{\mathrm{N}}^0)\geq \beta\mathrm{C}$ holds
 for any $P_0\in\mathbb{S}_{+}^n$.
Since $\mathrm{C}$ is arbitrarily chosen, the assertion follows as claimed.
The same is true of the statement $(ii)$ as the contraposition of $(i)$.
\hfill$\square$

There corresponds an analogy for $\liminf_{k\to \infty} \mathrm{Tr}(P_k)$ as we will present below. We omit the proof since it is similar to the proof of
Lemma~\ref{proposition:limsup-zero-one-law}.

\begin{lemma}\label{proposition:liminf-zero-one-law} Consider the deterministic evolution  (\ref{deterministic}).
\begin{itemize}
\item[(i)] If  there exists an initial condition
$P_0=\Sigma$ such that $\liminf\limits_{k\rightarrow \infty}\mathrm{Tr}(P_k)=\infty$, then\\
$\liminf\limits_{k\rightarrow \infty}\mathrm{Tr}(P_k)=\infty$
for all  $P_0\in\mathbb{S}_+^n$;

\item[(ii)] If there exists an initial condition
$P_0=\Sigma$ such that $\liminf\limits_{k\rightarrow \infty}\mathrm{Tr}(P_k)<\infty$, then\\
$\liminf\limits_{k\rightarrow \infty}\mathrm{Tr}(P_k)<\infty$
for all $P_0\in\mathbb{S}_+^n$.
\end{itemize}
\end{lemma}

The definition of tail events is as follows:
\begin{definition}
Let $\{\xi_k\}_{k\in\mathbb{N}}$ be a sequence of  random variables,
and  ${\tilde{\mathcal{F}}_k}\triangleq \sigma(\xi_k, \xi_{k+1},\ldots)$ be the smallest
$\sigma$--algebra generated by $\xi_k, \xi_{k+1},\ldots$ Then,
$\mathcal{T}(\{\xi_k\}_{k\in\mathbb{N}})\triangleq \cap_{j=1}^\infty {\tilde{\mathcal{F}}_j}$ is called
the tail algebra of $\{\xi_k\}_{k\in\mathbb{N}}$. If
$\mathcal{A}\in\mathcal{T}(\{\xi_k\}_{k\in\mathbb{N}})$, then $\mathcal{A}$ is said to be a tail event of $\{\xi_k\}_{k\in\mathbb{N}}$.
\end{definition}

We still need to recall  the concept  of {\it strong mixing}, which  was {first introduced} in~\cite{rosenblatt1956central}, and then
the zero--one law for strong mixing random processes established in \cite{arato1973twenty}. Note that $\ast$-mixing implies strong mixing \cite{bradley2005basic}.
\begin{definition}\label{def:strong-mixing}
The sequence of random variables $\{\xi_k\}_{k\in\mathbb{N}}$ on a probability space $(\mathscr{S},\mathcal{S},\mu)$ is said to be strong mixing if
$$\alpha(n)\triangleq \sup |\mu(\mathcal{A}\cap \mathcal{B})- \mu(\mathcal{A})\mu(\mathcal{B})|\to 0,~~\hbox{as}~n\to\infty,$$
where the supremum is taken over all $\mathcal{A}\in
\sigma(\xi_1,\ldots,\xi_k)$, $\mathcal{B}\in\sigma(\xi_{k+n},\xi_{k+n+1},
\ldots)$ and $k \in\mathbb{N}$.
\end{definition}

The following lemma holds.
\begin{lemma}
\label{lemma:Kolmogorov-zero-one-law} [Zero--One Law for Strong Mixing (Theorem 2.3, ~\cite{arato1973twenty})]
Let $\{\xi_k\}_{k\in\mathbb{N}}$ be a sequence of strong mixing random variables on a probability space $(\mathscr{S},\mathcal{S},\mu)$. Let
 $\mathcal{T}(\{{\xi_k}\}_{k\in\mathbb{N}})$ be the tail algebra of $\{\xi_k\}_{k\in\mathbb{N}}$. Then for any
$\mathcal{A}\in\mathcal{T}(\{\xi_k\}_{k\in\mathbb{N}})$, there holds $\mu(\mathcal{A})=1$ or $0$.
\end{lemma}

\begin{remark}
The concepts of $\ast$-mixing and strong mixing introduced in Definitions \ref{def:mixing} and \ref{def:strong-mixing}, originated in \cite{blum1963strong} and  \cite{rosenblatt1956central}, respectively, are imposed under different measures of dependence between past and future along  the random sequence. The $\ast$-mixing is equipped with a tighter measure, and therefore implies the strong mixing.  For a detailed introduction of their relations  we hereby refer to  \cite{bradley2005basic}, in which the $\ast$-mixing corresponds to $\psi$-mixing, and the strong mixing corresponds to $\alpha$-mixing.
\end{remark}

\subsection{Proof of Theorem \ref{thm:zero-one-law}}
We only focus on assertion $(i)$ for the event
$\mathcal{E}\triangleq\{\omega:\limsup\limits_{k\rightarrow \infty}\mathrm{Tr}(P_k)(\omega)<\infty\}$, as the conclusion for  $(ii)$  can be proved using the same argument.

Since $\mathrm{Tr}(P_k)$ is $\mathcal{F}$--measurable for all $k\in\mathbb{N}$,
so is $\limsup_{k\to \infty}\mathrm{Tr}(P_k)$. Then $\mathcal{E}\in\mathcal{F}$.
Fix a positive integer $n$ and a deterministic sequence $\{\tilde z_k\}_{k\in\mathbb{N}}$ with each $\tilde z_k$ taking value from $\{0,1\}$.
We define a sequence
$\{z_k\}_{k\in\mathbb{N}}$, where $z_k\in\{0,1\}$, such
that $z_k=\tilde z_{k+n}$.
Accordingly, we define a sequence of matrices
$\{\tilde P_k\}_{k\in\mathbb{N}}$ as the estimation error covariances
of the Kalman filter along $\{\tilde z_{k}\}_{k\in\mathbb{N}}$, where initial point is denoted by $\tilde P_0\in\mathbb{S}_+^n$; and define $\{ P_k\}_{k\in\mathbb{N}}$ along $\{z_{k}\}_{k\in\mathbb{N}}$, where initial point
is denoted by $P_0\in\mathbb{S}_+^n$. The rest of the proof consists of two aspects:
\begin{itemize}
\item[$(a)$]Suppose that $\limsup\limits_{k\to \infty}\mathrm{Tr}(\tilde P_k)=\infty$ holds with a given initial covariance $\tilde P_0$.
When $P_0=\tilde P_{n}$, we have $\limsup\limits_{k\rightarrow \infty}\mathrm{Tr}(P_k)=\limsup\limits_{k\rightarrow \infty}
\mathrm{Tr}(\tilde P_{k+n})=\infty$. It follows from $(i)$ of
Lemma~\ref{proposition:limsup-zero-one-law} that
$\limsup\limits_{k\rightarrow \infty}\mathrm{Tr}(P_k)=\infty$
holds for any $P_0\in\mathbb{S}_+^n$.

\item[$(b)$] Suppose that $\limsup\limits_{k\rightarrow \infty}\mathrm{Tr}(\tilde P_k)<\infty$ holds along $\{\tilde z_k\}_{k\in\mathbb{N}}$.
 Then we have $\limsup\limits_{k\rightarrow \infty}\mathrm{Tr}( P_k)=
\limsup\limits_{k\rightarrow \infty}\mathrm{Tr}(\tilde P_{k+n})<\infty$ when $P_0=\tilde P_{n}$. It follows from $(ii)$ of
Lemma~\ref{proposition:limsup-zero-one-law} that
$\limsup\limits_{k\rightarrow \infty}\mathrm{Tr}(P_k)<\infty$
holds for any $P_0\in\mathbb{S}_+^n$.
\end{itemize}
Define
$$\mathbb{S}_0\triangleq \big\{X: X=\phi_{\tilde z_n}\circ\cdots\circ\phi_{\tilde z_1}(\Sigma)\big\},$$
where $\tilde z_1,\ldots,\tilde z_n\in\{0,1\}$, and $\phi_i$ equals to the mapping $h$
when $i=0$ and $g$ when $i=1$.
It is straightforward that $\mathbb{S}_0$ is bounded (it is a finite set) and
$\mathbb{S}_0\subseteq\mathbb{S}_+^n$, therefore showing that
whether $\limsup\limits_{k\rightarrow \infty}\mathrm{Tr}(\tilde P_k)<\infty$ holds or not does not depend on $\tilde z_1,\ldots,\tilde z_n$.
Since $\{\tilde z_k\}_{k\in\mathbb{N}}$ is arbitrarily chosen from $\Omega$, we conclude that the event
$\mathcal{E}$ and its compliment $\mathcal{E}^c$
are independent of $\sigma(\gamma_1,\ldots,\gamma_n)$.
Again since $n$ is arbitrarily taken,  $\mathcal{E}\in
\mathcal{T}(\{\gamma_k\}_{k\in\mathbb{N}})$.
The conclusion then follows
from Lemma~\ref{lemma:Kolmogorov-zero-one-law}.

\subsection{Proof of Theorem \ref{thm:zero-one-law-2}}
First of all, we establish an auxiliary lemma.
\begin{lemma}\label{prop:zero-one-law-2}
Suppose $\mathrm{I_o}=1$. Then for any constant $\mathrm{C}>\mathrm{Tr}(M)$, where $M$ is defined in~\eqref{def:M}, the events
$\{\omega:\limsup_{k\to \infty } \mathrm{Tr}(P_k)(\omega) < \mathrm{C}\}$
and $\{\omega:\liminf_{k\to \infty } \mathrm{Tr}(P_k)(\omega) < \mathrm{C}\}$ are tail events of $\{\gamma_k\}_{k\in\mathbb{N}}$.
\end{lemma}
{\it Proof.} First let us show the conclusion for the event $\mathcal{A}_{\mathrm{C}}\triangleq\{\omega:\limsup\limits_{k\rightarrow \infty}\mathrm{Tr}(P_k)(\omega)<\mathrm{C}\}$. As in the proof of Theorem~\ref{thm:zero-one-law}, we can readily show
$\mathcal{A}_{\mathrm{C}}\in\mathcal{F}$.
Fix a positive integer $n$, a real number $\mathrm{C}>0$ and a deterministic sequence $\{\tilde z_k\}_{k\in\mathbb{N}}$ with each $\tilde z_k$ taking value from $\{0,1\}$.
we define a sequence
$\{z_k\}_{k\in\mathbb{N}}$, where $z_k\in\{0,1\}$, such
that $z_k=\tilde z_{k+n}$.
Accordingly, define a sequence of matrices
$\{\tilde P_k\}_{k\in\mathbb{N}}$ as the estimation error covariances
along $\{\tilde z_{k}\}_{k\in\mathbb{N}}$ with an initial point denoted by $\tilde P_0\in\mathbb{S}_+^n$; and define $\{P_k\}_{k\in\mathbb{N}}$
along $\{z_{k}\}_{k\in\mathbb{N}}$ with an initial point denoted by
$P_0\in\mathbb{S}_+^n$. The rest of the proof consists of two aspects:

\begin{itemize}
\item[$(a)$] Suppose that $\limsup\limits_{k\rightarrow \infty}\mathrm{Tr}(\tilde P_k)<\mathrm{C}$ holds with a given initial point $\tilde P_0$. In light of  Lemma~\ref{lemma:appedix-1}, we conclude via reduction to absurdity that $\{\tilde z_k\}_{k\in\mathbb{N}}\in \{\omega: \omega\in \gamma_k=1,{i.o.}\}$.
When $P_0=\tilde P_{n}\triangleq \Sigma_1$, we have $\limsup\limits_{k\rightarrow \infty}\mathrm{Tr}( P_k)=\limsup\limits_{k\rightarrow \infty}
\mathrm{Tr}(\tilde P_{k+n})<\mathrm{C}$. Next we shall now show  $\limsup\limits_{k\rightarrow \infty}\mathrm{Tr}(P_k)<\mathrm{C}$ for any $P_0\in\mathbb{S}_+^n$. Choose any matrix $\Sigma_2\in\mathbb{S}_+^n$. We differentiate $P_k$ with different initial points $\Sigma_1$ and $\Sigma_2$ by using notations $P_k^{\Sigma_1}$ and $P_k^{\Sigma_2}$ respectively.
By Lemma~\ref{lemma:appendix-positive-definite}, there exists an integer
$t$ such that $P_t^{\Sigma_1},P_t^{\Sigma_2}\in\mathbb{S}_{++}^n$.
Since $\mathrm{I_o}=1$, Lemma~\ref{lemma:contraction-mapping} indicates that
the operator $g$ is strictly contractive in $(\mathbb{S}_{++}^n,\delta)$. Therefore,
$$
\delta(P_k^{\Sigma_1},P_k^{\Sigma_2})\leq  q^{\sum_{i=t}^{k}z_i}
\delta(P_t^{\Sigma_1},P_t^{\Sigma_2})
$$
holds {for~all}~$k\geq t$, where $ q\in(0,1)$ is a constant that only depends on $A,~C,~Q,~R$.
As $k\to\infty$, we have $\sum_{i=t}^{k}z_i\to \infty$ and
consequently $\delta(P_k^{\Sigma_1},P_k^{\Sigma_2})\to 0$.
Thus, $P_k^{\Sigma_2}\to P_k^{\Sigma_1}$ due to the fact that
 $(\mathbb{S}_{++}^n,\delta)$ is a complete metric space.
Since $\mathrm{C}$ is arbitrarily chosen, $\limsup\limits_{k\rightarrow \infty}\mathrm{Tr}(P_k)<\mathrm{C}$ holds for any $P_0\in\mathbb{S}_+^n$.

\item[$(b)$]On the other hand, we suppose that $\limsup\limits_{k\rightarrow \infty}\mathrm{Tr}(\tilde P_k)\geq\mathrm{C}$ holds with a given initial covariance $\tilde P_0\in\mathbb{S}_+^n$. We discuss in all cases: $\limsup\limits_{k\rightarrow \infty}\mathrm{Tr}(\tilde P_k)$ is bounded by a larger constant $\tilde {\mathrm{C}}>{\mathrm{C}}$ or
    unbounded. For the first case, by using the same argument as in $(i)$,  $ \limsup\limits_{k\rightarrow \infty}\mathrm{Tr}(P_k)<\tilde{\mathrm{C}}$ holds for any $P_0\in\mathbb{S}_+^n$. For the other case, it follows form the proof of Theorem~\ref{thm:zero-one-law} that $\limsup\limits_{k\rightarrow \infty}\mathrm{Tr}(P_k)=\infty$ holds for any
    $P_0\in\mathbb{S}_+^n$.
\end{itemize}
Define
$$\mathbb{S}_0\triangleq \big\{X: X=\phi_{\tilde z_n}\circ\cdots\circ\phi_{\tilde z_1}(\Sigma)\big\},$$
where $\tilde z_1,\ldots,\tilde z_n\in\{0,1\}$, and $\phi_i$ equals to the mapping $h$
when $i=0$ and $g$ when $i=1$.
It is straightforward that $\mathbb{S}_0$ is bounded (it is a finite set) and
$\mathbb{S}_0\subseteq\mathbb{S}_+^n$. In view of the arguments in $(i)$ and $(ii)$, we obtain that
whether $\limsup\limits_{k\rightarrow \infty}\mathrm{Tr}(\tilde P_k)<\mathrm{C}$ holds or not does not depend on $\tilde z_1,\ldots,\tilde z_n$.
Since $\{\tilde z_k\}_{k\in\mathbb{N}}$ is arbitrarily chosen, the event $\mathcal{A}_{\mathrm C}$ and its compliment $(\mathcal{A}_{\mathrm C})^c$
are independent of $\sigma(\gamma_1,\ldots,\gamma_n)$. Again since $n$ and $\mathrm{C}$ are arbitrarily taken,
$\mathcal{A}_{\mathrm C}\in \mathcal{T}(\{\gamma_k\}_{k\in\mathbb{N}})$ holds for all $\mathrm{C}>0$.

It remains to show the assertion for the event $\mathcal{E}_{\mathrm C}\triangleq \{\omega:\liminf\limits_{k\rightarrow \infty}\mathrm{Tr}(P_k)(\omega)<\mathrm{C}\}$.
On one hand, by reduction to absurdity it is true for any $\mathrm{C}> \mathrm{Tr}(M)$ that
$$\mathcal{E}_{\mathrm C}
\subseteq \{\omega: \omega\in \gamma_k=1~{i.o.}\}.$$ On the other hand, from Lemma~\ref{lemma:suboptima-linear-estimator},
\begin{align}
\{\omega: \omega\in \gamma_k=1,{i.o.}\}&
\subseteq\{\omega:\liminf\limits_{k\rightarrow \infty}\mathrm{Tr}(P_k)(\omega)<\mathrm{Tr}(M)\}\subseteq
\mathcal{E}_{\mathrm C},
\end{align}
where the second ``$\subseteq$'' holds since $\mathrm{C}> \mathrm{Tr}(M)$.
In summary, we have \begin{equation}\label{eqn:relation-set}
\mathcal{E}_{\mathrm C}=\{\omega: \omega\in \gamma_k=1 ~{i.o.}\}.\end{equation}
Then $\mathcal{E}_{\mathrm C}\in\mathcal{T}(\{\gamma_k\}_{k\in\mathbb{N}})$ as one realizes that the latter event in~\eqref{eqn:relation-set} is a tail event.
\hfill$\square$

We are now in a place to complete the proof of Theorem \ref{thm:zero-one-law-2}. We only focus on the statement for absolutely upper a.s. stability, since that for absolutely lower a.s. stability can be analogously proved.
Define
$$\mathcal{A}_{x}\triangleq\{\omega:\limsup\limits_{k\rightarrow \infty}\mathrm{Tr}(P_k)(\omega)<x\},~~~x>0.$$
It is clear that $\mathcal{A}_{\lfloor x\rfloor}\mathcal\subseteq\mathcal{A}_{x}\subseteq\mathcal{A}_{\lceil x\rceil}$
holds for all $x>0$,
which eventually results in
$$\bigcup_{\mathrm{C}\in(0,\infty)}\mathcal{A}_\mathrm{C}=
\bigcup_{\mathrm{C}\in\mathbb{N}}\mathcal{A}_\mathrm{C}.$$
Since $\mathcal{A}_\mathrm{C}\in\mathcal{T}(\{\gamma_k\}_{k\in\mathbb{N}})$ for any $\mathrm{C}> \mathrm{Tr}({M})$ by Lemma~\ref{prop:zero-one-law-2},
$$\bigcup_{\mathrm{C}\in(0,\infty)}\mathcal{A}_\mathrm{C}=
\bigcup_{\mathrm{C}\in\mathbb{N},\atop \mathrm{C}> \mathrm{Tr}({M})}\mathcal{A}_\mathrm{C}\in \mathcal{T}(\{\gamma_k\}_{k\in\mathbb{N}}).$$
Finally, the  conclusion follows
from Lemma~\ref{lemma:Kolmogorov-zero-one-law}.

\section{Almost Sure Stability Conditions}\label{section:stability-conditions}

{In the last section, we have shown that whether the considered
Kalman filter is a.s. stable or not can be interpreted
by a zero--one law.  In this section, we are devoted to studying the relationship between the packet rate and these stability notions. We first present some sufficient/necessary stability conditions for general LTI systems. Then we continue to show that one--step observable systems admit tighter results, with
necessary and sufficient conditions derived for upper and lower a.s. stabilities, respectively.
Finally, for the so--called non--degenerate systems, we  give a necessary and sufficient upper a.s. stability condition.
%

Denote $\mathbb{E}[\gamma_k]\triangleq p_k$. To make the analysis concise,
we require the following
assumption
\begin{enumerate}
\item[(A3)]\label{asmpt:assumpt-A}
\textit{$\{ p_k\}_{k\in {\mathbb{N}}}$
is a monotonic sequence.}
\end{enumerate}
It is not difficult to find that, if
(A3) is not satisfied, all results are still tractable under the current analysis
but in more complex forms.
We  choose (A3) to be our standing assumption  in the rest of this section.

\subsection{Main Results}

\subsubsection{General Stability Conditions}
First we
give sufficient conditions for (absolute) lower a.s. stability and
lower a.s. instability. Recall that $\mathrm{I_o}$ is the observability
index defined in~Definition~\ref{def:observability-index}.
\begin{theorem}\label{thm:liminf-as-bouned} Let Assumptions (A1)--(A3) hold.
\begin{itemize}
\item[(i)]If
{$\sum\limits_{k=1}^{\infty}( p_k)^{\mathrm{I_o}}=\infty$}, then
the considered filtering system is absolutely lower a.s. stable for any $P_0\in\mathbb{S}_+^n$.

\item[(ii)]If $\sum\limits_{k=1}^{\infty}p_k<\infty$, then the considered filtering system is lower a.s. unstable for any $P_0\in\mathbb{S}_+^n$.
\end{itemize}
\end{theorem}

The following theorem presents a necessary condition for upper a.s. stability.
\begin{theorem}\label{thm:limsup-as-bouned}
Let Assumptions (A1)--(A3) hold.  If the considering filtering system is upper a.s. stable, then
there exists a constant $\mathrm{I}\in\mathbb{N}$ such that $\sum\limits_{k=1}^{\infty}(1- p_k)^{\mathrm{I}}<\infty$.
\end{theorem}

The proofs of Theorems~\ref{thm:liminf-as-bouned} and~\ref{thm:limsup-as-bouned} rely on Borel-Cantelli lemmas with respect to $*$-mixing.

\subsubsection{One--step Observable Systems}\label{subsection:one-step-obs-sys}

As a special case, one--step observable systems have $\mathrm{I_o}=1$. The following theorem
provides necessary and sufficient conditions for (absolutely) lower a.s. stability.
\begin{theorem}\label{thm:lower-as-stability-one-step} Let Assumptions (A1)--(A3) hold.
Suppose $\mathrm{I_o}=1$.
For any $P_0\in\mathbb{S}_+^n$, the following conditions are equivalent:
\begin{itemize}
\item[(i)]The considered filtering system is absolutely lower a.s. stable;

 \item[(ii)]The considered filtering system is lower a.s. stable;

 \item[(iii)]There holds that $\sum\limits_{k=1}^{\infty}p_k=\infty$.
\end{itemize}
\end{theorem}

In the following, we also  present necessary and sufficient conditions
for (absolutely) upper a.s. stability.
\begin{theorem}\label{thm:limsup-as-bouned-first-order}
Let Assumptions (A1)--(A3) hold.
Suppose $\mathrm{I_o}=1$.  For any $P_0\in\mathbb{S}_+^n$,
the following statements are equivalent:
\begin{itemize}
\item[(i)]The considered filtering system is absolutely upper a.s. stable;

 \item[(ii)]The considered filtering system is upper a.s. stable;

 \item[(iii)]There exists a constant $\mathrm{I}\in\mathbb{N}$ such that $\sum\limits_{k=1}^{\infty}(1- p_k)^{\mathrm{I}}<\infty$.
\end{itemize}
\end{theorem}

Theorems~\ref{thm:lower-as-stability-one-step}~and~\ref{thm:limsup-as-bouned-first-order}
are proved, partially relying on the fact that, as long as the considered Kalman filter successfully receives a packet, its instantaneous error covariance is bounded from above. The detailed proofs have been
put in Section~\ref{subsection:proof-section4}.
\subsubsection{Non--degenerate Systems}
For general LTI systems with $\mathrm{I_o}\geq 2$, it is challenging to
find conditions guaranteeing (absolutely) upper a.s. stability, since $P_k$ does not necessarily decrease when
packets are intermittently received. However, an exception is a class of so--called non--degenerate systems. We first introduce the definition of
non--degenerate systems, which is taken from~\cite{yilin12criticalvalue,you2011mean}, and then present the probabilistic stability guarantor of $\sup_{k\geq n} \mathrm{Tr}(P_k)$ for this kind of systems. Note that the requirement of non--degeneracy is indispensable because it enables us to
bound $\mathrm{Tr}(P_k)$ when intermittent receptions of measurements happen.
\begin{definition}
Consider a system $(C,A)$ in diagonal
standard form, i.e., $A=\mathrm{diag}
(\lambda_1,\ldots,\lambda_n)$ and
$C=[C_1,\ldots,C_n]$. An quasi--equiblock
of the system is defined as a subsystem
$(C_\mathcal{I},A_\mathcal{I})$, where
$\mathcal{I}\triangleq \{l_1,\ldots,l_i\}
\subset\{1.\ldots,n\}$, such that
$A_\mathcal{I}=\mathrm{diag}(\lambda_{l_1},\ldots,\lambda_{l_i})$
with $|\lambda_{l_1}|=\cdots=|\lambda_{l_i}|$ and
$C_\mathcal{I}=[C_{l_1},\ldots,C_{l_i}]$.
\end{definition}
\begin{definition}
A diagonalizable system $(C,A)$ is non--degenerate if
every quasi--equiblock of the system is one--step observable. Conversely, it is
degenerate if it has at least one quasi--equiblock that is
not one--step observable.
\end{definition}

The following result holds.

\begin{theorem}\label{thm:liminf-as-bouned-nondegenerate}
Let Assumptions {(A1)--(A3)} hold.
Suppose the system $(C,A)$ is non--degenerate.
For any $P_0\in\mathbb{S}_+^n$,
the following statements are equivalent:
\begin{itemize}
\item[(i)] The considered filtering system is absolutely upper a.s. stable;

 \item[(ii)] The considered filtering system is upper a.s. stable;

 \item[(iii)] There exists a constant $\mathrm{I}\in\mathbb{N}$ such that $\sum\limits_{k=1}^{\infty}(1- p_k)^{\mathrm{I}}<\infty$.
\end{itemize}
\end{theorem}

\begin{remark}
The necessary and sufficient conditions in
Theorem~\ref{thm:liminf-as-bouned-nondegenerate} suggest that,
when $(C,A)$ is non--degenerate, the absolutely upper a.s. stability
also follows a zero--one law.
\end{remark}

\subsection{Supporting Lemmas}
This subsection presents supporting lemmas and auxiliary definitions for the proofs of the
main results.
The following two lemmas concern with sequences of real numbers.
The first one is well known and its proof can be found in~\cite{rudin1987real}.
\begin{lemma}
Suppose that $\{ a_k\}_{k\in {\mathbb{N}}}$ is a sequence of real numbers with
$ a_k\in [0,1)$. Then $\sum_{k=1}^\infty  a_k=\infty$ holds if and only if $\prod_{k=1}^\infty (1- a_k)=0$.
\end{lemma}

\begin{lemma}\label{lemma:monotonic-sequence}
Suppose that $\{ a_k\}_{k\in {\mathbb{N}}}$ is a monotonic sequence of real numbers with
$ a_k\in [0,\infty)$.
Then, for any $l\geq 2$, $\sum\limits_{i=0}^\infty \prod_{k=il+1}^{(i+1)l} a_k=\infty$ holds
if and only if $\sum\limits_{k=1}^\infty ( a_k)^l=\infty$.
\end{lemma}
{\it Proof.}
Without loss of generality, we assume that $\{ a_k\}_{k\in {\mathbb{N}}}$ is monotonically
decreasing, for a monotonically increasing sequence can be treated in a similar manner.
 For simplicity, let $s_j\triangleq \sum\limits_{i=1}^\infty \prod_{k=(i-1)n+j}^{in+j-1} a_k$ for $j\in {\mathbb{N}}$.
If $s_{1}=\infty$, observing that $s_1\geq s_2\geq \cdots \geq s_{n}\geq s_{n+1}$, and
that $s_{n+1}=s_{1}-\prod_{k=1}^{n} a_k$,
we have $s_j=\infty$.
Therefore,
$$
\sum_{j=1}^{n} s_j
\leq \sum_{k=1}^\infty ( a_k)^n,
$$
implying $\sum_{k=1}^\infty ( a_k)^n=\infty$.
To prove the sufficiency, note that
$$
n s_1\geq \sum_{j=1}^{n} s_j\geq \sum_{k=n}^\infty ( a_k)^n.
$$
Since $n$ is finite, the desired conclusion follows.
\hfill$\square$

The following lemma is the first Borel--Cantelli lemma from probability theory. For more details, please refer to~\cite{durrett2010probability}.
\begin{lemma}\label{lemma:borel-cantelli-lemma} [First Borel--Cantelli lemma]
Let $(\mathscr{S}, \mathcal{S}, \mu)$ be a probability space. Suppose $\{\mathcal{A}_j\}_{j\in\mathbb{N}}$
is a sequence of events, where
$\mathcal{A}_j\in \mathcal{S}~\hbox{for~all}~j\in{\mathbb{N}}$.
If $\sum\limits_{i=1}^{\infty}\mu(\mathcal{A}_i)<\infty$, then $\mu\left
(\mathcal{A}_i{~i.o.}\right)=0$.
\end{lemma}

The definition of
$*$-mixing for a sequence of events, and the corresponding second Borel--Cantelli lemma are as follows.

\begin{definition}
A sequence of events $\{\mathcal{A}_j\}_{j\in\mathbb{N}}$ is said to be
$*$-mixing if $\{1_{\mathcal{A}_j}\}_{j\in\mathbb{N}}$ is
$*$-mixing.
\end{definition}
\begin{lemma}\label{lemma:borel-cantelli-lemma-2} [Second Borel--Cantelli lemma under $*$--Mixing (Lemma~6,~\cite{blum1963strong})]
Let $\{\mathcal{A}_j\}_{j\in\mathbb{N}}$
be a sequence of  $*$-mixing events on a probability space $(\mathscr{S}, \mathcal{S}, \mu)$.
Then  $\mu\left(\mathcal{A}_i~{i.o.}\right)=1$ if $\sum\limits_{i=1}^{\infty}\mu(\mathcal{A}_i)=\infty$.
\end{lemma}

We define the following two quantities to evaluate the minimum
and maximum lengths of consecutive packet drops that make the error covariance
exceed a given threshold. With the help of the two
quantities, we develop a sufficient condition for that $\limsup\limits_{k\rightarrow \infty}\mathrm{Tr}(P_k)$ exceeds a given threshold almost surely. For a one--step observable system (i.e., $\mathrm{I_o}=1$), as long as
one packet is received, $P_k\leq {M}$ holds by Lemma~\ref{lemma:suboptima-linear-estimator}, enabling us
to develop {necessary conditions} for such a system.

Let us
define two quantities $\overline{{\mathrm{I}}}(\mathrm{C})$ and
$\underline{{\mathrm{I}}}(\mathrm{C})$ as follow: for a given real number $\mathrm{C}\geq \mathrm{Tr}(M)$, put
\begin{eqnarray}
\overline{{\mathrm{I}}}(\mathrm{C})&\triangleq&\min\left\{k\in\mathbb{N}: \mathrm{Tr}\big(h^{k}(M)\big)> \mathrm{C}\right\},\label{eqn:def-overlineIM}\\
\underline{{\mathrm{I}}}(\mathrm{C})&\triangleq&\min\left\{k\in \mathbb{N}: \mathrm{Tr}\big(h^{k}(\overline P)\big)> \mathrm{C}\right\}.\label{eqn:def-underlineIM}
\end{eqnarray}
Similar definitions for $\overline{{\mathrm{I}}}(\mathrm{C})$ and
$\underline{{\mathrm{I}}}(\mathrm{C})$ primarily appeared in~\cite{shi-tac10}, where the quantities were
used to derive upper and lower bounds on $\mathbb{P}\left(P_{k|k}\leq P_*\right)$ for some $P_*\in\mathbb{S}_+^n$.
Different from~\cite{shi-tac10},
in this paper, we will use
these two quantities to characterize the relationships between
the packet rate and
various stability notations in Definition~\ref{def:as-statbility}.

The following lemma says that, for any $\mathrm{C}\geq \mathrm{Tr}({M})$, both $\overline{{\mathrm{I}}}(\mathrm{C})$ and
$\underline{{\mathrm{I}}}(\mathrm{C})$ are bounded.
\begin{lemma}\label{lemma:finite-window-times}
Suppose $A$ is unstable. {Then, there} holds
$\overline{{\mathrm{I}}}(\mathrm{C})\leq \underline{{\mathrm{I}}}(\mathrm{C})<\infty $~for~all~$\mathrm{C}\geq \mathrm{Tr}({M}).$
\end{lemma}
\textit{Proof:~}
First of all, it is evident from Lemma~\ref{lemma:appendix-g-h} that
$\overline{{\mathrm{I}}}(\mathrm{C})\leq \underline{{\mathrm{I}}}(\mathrm{C})$.
Since $\overline{P}<M$ by Lemma~\ref{lemma:suboptima-linear-estimator},
to show that $\overline{{\mathrm{I}}}(\mathrm{C})$ and $\underline{{\mathrm{I}}}(\mathrm{C})$ are finite
for any $\mathrm{C}\geq \mathrm{Tr}({M})$, it suffices to show that there exists an integer $k\in \mathbb{N}$ implying $\mathrm{Tr}\left(h^k(\overline{P})\right)>\mathrm{C}$.
By Lemma~\ref{lemma:appedix-1}, there always exists an $  a >0$ such that $\mathrm {Tr}\left(h^k(X)\right)\geq  a |\lambda_1(A)|^{2k}$.
Therefore, when taking
$$k\geq \left\lceil\frac{\mathrm{Log} \mathrm{C}-\mathrm{Log} a}{2\mathrm{Log}
|\lambda_1(A)|}\right\rceil+1,$$
we have  $ a |\lambda_1(A)|^{2k}> \mathrm{C}$, which completes the proof.
\hfill $\square$

\begin{lemma}\label{prop:limsup-as-bouned-first-order}
Suppose that $\mathrm{I_o}=1$. Consider a real number $\mathrm{C}\geq \mathrm{Tr}({M})$.
If $\sum\limits_{k=1}^{\infty}(1- p_k)^{\overline{{\mathrm{I}}}(\mathrm{C})}<\infty$,
then $\mathbb{P}\big(\limsup\limits_{k\rightarrow \infty}\mathrm{Tr}(P_k)\leq
\mathrm{C}\big)=1$ holds for all $P_0\in\mathbb{S}_+^n$.
\end{lemma}
\textit{Proof:~}{
Noticing \begin{align*}
&\prod_{k=i}^{\overline{{\mathrm{I}}}(\mathrm{C})+i-1}
(1- p_k)\leq \max\bigg\{(1- p_k)^{\overline{{\mathrm{I}}}(\mathrm{C})}:i\leq k \leq \overline{{\mathrm{I}}}(\mathrm{C})+i-1\bigg\}\leq
\sum_{k=i}^{\overline{{\mathrm{I}}}(\mathrm{C})+i-1}
(1- p_k)^{\overline{{\mathrm{I}}}(\mathrm{C})},
\end{align*}
we have
\begin{align}\label{r1}
\sum_{i=1}^\infty \prod_{k=i}^{\overline{{\mathrm{I}}}(\mathrm{C})+i-1}
(1- p_k) &\leq \sum_{i=1}^\infty
\sum_{k=i}^{\overline{{\mathrm{I}}}(\mathrm{C})+i-1}
(1- p_k)^{\overline{{\mathrm{I}}}(\mathrm{C})}.
\end{align}
Since $\overline{{\mathrm{I}}}(\mathrm{C})$ is a finite number,  each term $(1- p_j)^{\overline{{\mathrm{I}}}(\mathrm{C})}$ appears at most   $\overline{{\mathrm{I}}}(\mathrm{C})$ times in the summation of the right-hand side of  (\ref{r1}) for any $j\geq 1$. As a result,
\begin{align*}
 \sum_{i=1}^\infty
\sum_{k=i}^{\overline{{\mathrm{I}}}(\mathrm{C})+i-1}
(1- p_k)^{\overline{{\mathrm{I}}}(\mathrm{C})} \leq\overline{{\mathrm{I}}}(\mathrm{C})\sum_{k=1}^{\infty}(1- p_k)^{\overline{{\mathrm{I}}}(\mathrm{C})}.
\end{align*}
This leads to
\begin{align*}
\sum_{i=1}^\infty \prod_{k=i}^{\overline{{\mathrm{I}}}(\mathrm{C})+i-1}
(1- p_k)
\leq\overline{{\mathrm{I}}}(\mathrm{C})\sum_{k=1}^{\infty}(1- p_k)^{\overline{{\mathrm{I}}}(\mathrm{C})}
<\infty,
\end{align*}
from the standing hypothesis.

Next, from Lemma~\ref{lemma:borel-cantelli-lemma} and the definition of $\overline{{\mathrm{I}}}(\mathrm{C})$, we see
$\mathbb{P}\left(\mathrm{Tr}(P_k) > \mathrm{C}~{i.o.}
\right)=0$ for any $P_0\in\mathbb{S}_+^n$ since the event $\mathrm{Tr}(P_k) > \mathrm{C}$ requires that at least $\overline{{\mathrm{I}}}(\mathrm{C})$ consecutive drops happen before the time $k$. This completes the proof. }
\hfill$\square$

\subsection{Proofs of Statements}\label{subsection:proof-section4}
\subsubsection{Proof of Theorem~\ref{thm:liminf-as-bouned}}
If $\sum\limits_{k=1}^{\infty}( p_k)^{\mathrm{I_o}}=\infty$, by Lemma~\ref{lemma:monotonic-sequence},
one obtains \vspace{1mm}$\sum\limits_{i=0}^\infty \prod_{k=i\mathrm{I_o}+1}^{(i+1)\mathrm{I_o}} p_k=\infty$.

Define $$\mathcal{A}_j\triangleq\bigg\{\omega: \prod_{l=j\mathrm{I_o}+1}^
{(j+1)\mathrm{I_o}}
\gamma_l(\omega)=1,\omega\in\Omega\bigg\},~~~j\in\mathbb{N}.$$
Since $\{\gamma_k\}_{k\in\mathbb{N}}$ is $*$-mixing and
$\mathrm{I_o}\leq n$, the sequence
$\{\mathcal{A}_j\}_{j\in\mathbb{N}}$ of events induced by
$\{\gamma_k\}_{k\in\mathbb{N}}$ is $*$-mixing by definition.
By Lemmas~\ref{lemma:suboptima-linear-estimator}~and~\ref{lemma:borel-cantelli-lemma-2}, we have
$$\mathbb{P}\left(P_k\leq{M}~{i.o.}\right)\geq \mathbb{P}(\mathcal{A}_j~{i.o.})=1,$$ where the first assertion follows.

If $\sum_{k=1}^\infty p_k<\infty$,  $\mathbb{P}\left(\gamma_k=1~{i.o.}\right)=0$ holds by  Lemma~\ref{lemma:borel-cantelli-lemma}. Then, by Lemma~\ref{lemma:appedix-1}, there holds
$$\mathbb{P}\Big(\liminf_{k\to \infty } \mathrm{Tr}(P_k) < \infty\Big)=0,$$
which completes the proof.

\subsubsection{Proof of Theorem~\ref{thm:limsup-as-bouned}}
We shall prove the contraposition of the theorem, viz. that,
if $\sum\limits_{k=1}^\infty(1-p_k)^\mathrm{I}=\infty$ for any $\mathrm{I}\in\mathbb{N}$, then the considered filtering system is upper
a.s. unstable.
To this end, fix any constant $\mathrm{C}\geq \mathrm{Tr}(M)$ and a realization $\omega\in\Omega$ of $\{\gamma_k\}_{k\in\mathbb{N}}$. By
the definition of $\underline{\mathrm I}(\mathrm{C})$ in~\eqref{eqn:def-underlineIM} and
Lemma~\ref{lemma:finite-window-times}, we have $\mathrm{Tr}\left(h^{\underline{\mathrm{I}}(\mathrm{C})}(\overline{P})\right)
>\mathrm{C}$ and $\underline{\mathrm{I}}(\mathrm{C})<\infty$.
Then, from {the continuity} of
the matrix trace and $h$ operators, there always exists a constant $\beta\in(0,1)$ such that
$\mathrm{Tr}\left( h^{\underline{{\mathrm{I}}}(\mathrm{C})}(\beta\overline{P})\right)>\mathrm{C}.$
Since $\lim_{k\rightarrow \infty}g^k(0)=\overline{P}$, there exists a sufficiently large $\mathrm{N}(\beta)$ that implies  $g^k(0)\geq \beta \overline{P}$ for all
$k\geq \mathrm{N}(\beta)$, see the proof of Lemma~\ref{lemma:appendix-positive-definite} at this statement.
By~\eqref{eqn:h-geq-g} in Lemma~\ref{lemma:appendix-g-h}, $P_k(\omega)\geq g^k(0)>\beta\overline P$ holds for all $k\geq \mathrm{N}(\beta)$.
This observation therefore leads to $\mathrm{Tr}\big(h^{\underline{{\mathrm{I}}}
(\mathrm{C})}({P}_k(\omega))\big)>\mathrm{C}$ for all $k\geq \mathrm{N}(\beta)$.
When taking all $\omega$'s within $\Omega$ into account, we have
\begin{equation}\label{eqn:set-relation}
\mathcal{E}_{\underline{{\mathrm{I}}}(\mathrm{C})}\subseteq(\mathcal{A}_{\mathrm C})^c,
\end{equation}
where $\mathcal{E}_{\underline{{\mathrm{I}}}(\mathrm{C})}\triangleq$$\{\omega: \underline{{\mathrm{I}}}(\mathrm{C})$~numbers of consecutive
packet dropouts occur}~{i.o.}$\}$ and $\mathcal{A}_{\mathrm C}\triangleq \{\omega:\limsup_{k\to\infty}{\mathrm{Tr}(P_k(\omega))}\leq\mathrm{C}\}$.
In addition, the hypothesis $\sum\limits_{k=1}^{\infty}(1- p_k)^{\underline{{\mathrm{I}}}(\mathrm{C})}=\infty$  implies
$$\sum\limits_{i=0}^\infty \prod_{k=i\underline{{\mathrm{I}}}(\mathrm{C})+1}
^{(i+1)\underline{{\mathrm{I}}}(\mathrm{C})}
(1- p_k)=\infty
$$ by Lemma~\ref{lemma:monotonic-sequence}.
Define
$$
\mathcal{B}_j=\bigg\{\omega:
\prod_{l=j\underline{{\mathrm{I}}}(\mathrm{C})+1}
^{(j+1)\underline{{\mathrm{I}}}(\mathrm{C})}
\big(1-\gamma_l(\omega)\big)=1,\omega\in\Omega\bigg\},~~~~j\in\mathbb{N}.
$$
Since $\{\gamma_k\}_{k\in\mathbb{N}}$ is $*$-mixing and
$\underline{{\mathrm{I}}}(\mathrm{C})<\infty$, the events
$\{\mathcal{B}_j\}_{j\in\mathbb{N}}$ induced by
$\{\gamma_k\}_{k\in\mathbb{N}}$ is $*$-mixing by definition.
By virtue of  Lemma~\ref{lemma:borel-cantelli-lemma-2}, it implies that
\begin{equation}\label{eqn:prob-one}
\mathbb{P}(\mathcal{E}_{\underline{{\mathrm{I}}}(\mathrm{C})})\geq \mathbb{P}(\mathcal{B}_i~{i.o.})=
1.
\end{equation}

Since $\mathrm{C}$ is arbitrarily chosen from the interval $[\mathrm{Tr}(M), \infty)$,
\begin{align*}
&\{\omega:\limsup_{k\to\infty}{\mathrm{Tr}(P_k(\omega))<}\infty\}= \bigcup_{\mathrm{C}\in[\mathrm{Tr}(M),\infty)}\mathcal{A}_{\mathrm C}
\subseteq \bigcup_{\mathrm{C}\in[\mathrm{Tr}(M),\infty)} (\mathcal{E}_{\underline{{\mathrm{I}}}(\mathrm{C})})^c
\subseteq\bigcup_{\underline{{\mathrm{I}}}(\mathrm{C})=1}^\infty (\mathcal{E}_{\underline{{\mathrm{I}}}(\mathrm{C})})^c,
\end{align*}
where the first ``$\subseteq$'' is from~\eqref{eqn:set-relation}
and the second one is due to $\underline{\mathrm{I}}(\mathrm{C})<\infty$. As a result,
\begin{align*}
\mathbb{P}(\limsup_{k\to\infty}{\mathrm{Tr}(P_k)<}\infty)
&\leq  \mathbb{P}\Big(\bigcup_{\underline{{\mathrm{I}}}(\mathrm{C})=1}^\infty (\mathcal{E}_{\underline{{\mathrm{I}}}(\mathrm{C})})^c\Big)\leq \sum_{\underline{{\mathrm{I}}}(\mathrm{C})=1}^\infty
\Big(1-\mathbb{P}(\mathcal{E}_{\underline{{\mathrm{I}}}(\mathrm{C})})\Big)= 0,
\end{align*}
in which the second inequality is due to subadditivity of measure $\mathbb P$ and the last equality is due to~\eqref{eqn:prob-one}.
This completes the proof.

\subsubsection{Proof of Theorem~\ref{thm:lower-as-stability-one-step}}
$(i)\Rightarrow(ii)$ is true from the definition in its own right.
Since $\mathrm{I_o}=1$ and the fact that
lower a.s. stability {follows the one--zero law}, $(ii)\Rightarrow(iii)$ and $(iii)\Rightarrow(i)$ hold by Theorem~\ref{thm:liminf-as-bouned}.

\subsubsection{Proof of Theorem~\ref{thm:limsup-as-bouned-first-order}}
Note that $(i)$ implies $(ii)$ by definition and $(ii)$ implies $(iii)$
by Theorem~\ref{thm:limsup-as-bouned}. It remains to show $(iii)\Rightarrow (i)$. Take a constant $\mathrm{C}$ such that $\mathrm{C}\geq \min\{\mathrm{Tr}(M),h^{\mathrm{I}}(M)\}$.
By~\eqref{eqn:def-overlineIM} and Lemma~\ref{lemma:finite-window-times}, we have $\mathrm{I}< \overline{{\mathrm{I}}}(\mathrm{C})<\infty $. Then $(iii)$ implies $\sum_{k=1}^\infty(1-p_k)^{\overline{{\mathrm{I}}}(\mathrm{C})}<\infty$.
According to Lemma~\ref{prop:limsup-as-bouned-first-order},
$\mathbb{P}\big(\limsup\limits_{k\rightarrow \infty}\mathrm{Tr}(P_k)\leq
\mathrm{C}\big)=1$ holds for all $P_0\in\mathbb{S}_+^n$, which completes the proof. 

\subsubsection{Proof of Theorem~\ref{thm:liminf-as-bouned-nondegenerate}}
Similar to the proof of Theorem~\ref{thm:limsup-as-bouned-first-order}, we only need to show $(iii)\Rightarrow (i)$. To this end,
we first define a sequence of stopping time $\{t_j\}_{j\in \mathbb{N}}$
as a sequence of packet arrival times
as follows:
\begin{align*}
t_1&\triangleq \min\{k:k\geq 1, \gamma_k=1\},\\
&\vdots \\
t_j&\triangleq \min\{k:k>t_{j-1}, \gamma_k=1\}.
\end{align*}
If $\max\{j:t_j\leq k\}\geq n$, it means that the estimator has
received no less than $n$ packets up to
time $k$. In this case, we define
\begin{align*}
\tau_{k,1}&\triangleq k-t_i\hbox{~~where~}i=\max\{j:t_j\leq k \}, \hbox{~and}\\
\tau_{k,j}&\triangleq t_{i-j+2}-t_{i-j+1},
\hbox{~for~} 2\leq j \leq n.
 \end{align*}
To get the desired result, we need the following lemma.
\begin{lemma}\label{lemma:lemma-yilin}
 If $\max\{j:t_j\leq k\}\geq n$ and the system is non--degenerate,
then the following inequality holds:
$$\mathrm{Tr}(P_{k+1})\leq  a_0 \prod_{j=1}^n (|\lambda_1(A)|+\epsilon)^{2\tau_{k,j}},$$
where ${  a_0}$ is a constant independent of $\tau_{k,j}$ and
$\epsilon$ can be arbitrarily small.
\end{lemma}
{\it Proof.}
The result can be readily established from Theorem 4 in~\cite{yilin12criticalvalue}
and the fact that $|\lambda_1(A)|\geq\cdots\geq |\lambda_n(A)|$.
\hfill$\square$

If there exists an $\mathrm{I}\in\mathbb{N}$ such that
$\sum_{k=0}^{\infty}(1- p_k)^{\mathrm{I}}<\infty$,  we can always find
a sufficiently large positive number $\mathrm{C}_\mathrm{I}$ satisfying
$\mathrm{C}_\mathrm{I}>   a_0(|\lambda_1(A)|+\epsilon)^{2(n+\mathrm{I}-2)}$
for a small $\epsilon>0$.
 Given any time $k\geq n+\mathrm{I}-1$, we compute
\begin{align}\label{eqn:prob-bound}
&\mathbb{P}\bigg(\mathrm{Tr}(P_{k+1})>\mathrm{C}_\mathrm{I}
\bigg)\nonumber\\
&\leq \mathbb{P}\bigg(\mathrm{Tr}(P_{k+1})>  a_0 (|\lambda_1(A)|+\epsilon)^{2(n+\mathrm{I}-2)}\bigg)\notag\\
&\leq
\mathbb{P}\bigg(
\hbox{less~than~}n\hbox{~packets~received~between}\hbox{~time~}k-n-\mathrm{I}+2
\hbox{~and~}k
\bigg)\notag\\
&\leq\sum_{j=0}^{n-1} \left(\begin{array}{cc}
n+\mathrm{I}-1
\\
j\end{array}\right)
\max\{p_{k-n-\mathrm{I}+2},p_{k}\}^j\left(1-\min\{p_{k-n-\mathrm{I}+2},p_{k}\}\right)^{n+\mathrm{I}-j-1} \notag\\
&\leq\sum_{j=0}^{n-1} \left(\begin{array}{cc}
n+\mathrm{I}-1
\\
j\end{array}\right)\left(1-\min\{p_{k-n-\mathrm{I}+2},p_{k}\}\right)^{\mathrm{I}}\notag\\
&\leq\sum_{j=0}^{n-1} \left(\begin{array}{cc}
n+\mathrm{I}-1
\\
j\end{array}\right)\left(1-p_{k-n-\mathrm{I}+2}\right)^{\mathrm{I}}+
\sum_{j=0}^{n-1} \left(\begin{array}{cc}
n+\mathrm{I}-1
\\
j\end{array}\right)\left(1-p_{k}\right)^{\mathrm{I}}
\end{align}
where the second inequality holds due to Lemma~\ref{lemma:lemma-yilin} and the observation that $\sum_{j=1}^n\tau_{k,j}\leq n+\mathrm{I}-2$ if and only if less than $n$ packets are received between
time $k-n-\mathrm{I}+2$ and $k$,
the second last inequality is from the monotonicity of $\{ p_k\}_{k\in {\mathbb{N}}}$, and $\left(\begin{array}{cc}
\cdot
\\
\cdot\end{array}\right)$ denotes a combination number.
Thus, \begin{align*}
&\sum_{k=1}^\infty\mathbb{P}\left(\mathrm{Tr}(P_{k})>\mathrm{C}_\mathrm{I}
\right)\nonumber\\
&=\sum_{k=1}^{n+\mathrm{I}-1}\mathbb{P}\left(\mathrm{Tr}(P_{k})>\mathrm{C}_\mathrm{I}
\right)+\sum_{k=n+\mathrm{I}}^{\infty}\mathbb{P}\left(\mathrm{Tr}(P_{k})>\mathrm{C}_\mathrm{I}
\right)\\
&\leq \sum_{k=1}^{n+\mathrm{I}-1}\mathbb{P}\left(\mathrm{Tr}(P_{k})>\mathrm{C}_\mathrm{I}
\right)+2\sum_{j=0}^{n-1} \left(\begin{array}{cc}
n+\mathrm{I}-1
\\
j\end{array}\right)\sum_{k=1}^{\infty}(1- p_k)^{\mathrm{I}}\\
&<\infty,
\end{align*}
where the first inequality follows from~\eqref{eqn:prob-bound}.
By Lemma~\ref{lemma:borel-cantelli-lemma},
$\mathbb{P}\left(\mathrm{Tr}(P_k)>\mathrm{C}_\mathrm{I}~{i.o.}\right)=0$
holds even for the set of events
$\{\omega:\mathrm{Tr}(P_k(\omega))>\mathrm{C}_\mathrm{I}\}_{k\in {\mathbb{N}}}$ that are not independent,
which completes the proof.

\section{Conclusions}\label{section:conclusion}
We have studied the stability, from the
probabilistic perspective, of Kalman filtering with random
packet dropouts.
The packet dropouts were modeled  by a $*$-mixing model,
 whereby
the occurrence of any two packet drops can be considered approximately
``independent'' as they are sufficiently far apart from each other.
We defined
(absolutely) upper and lower a.s. stabilities of the considered filtering systems.
We established a zero--one law of upper and lower a.s. stabilities for {general} LTI systems, which makes the upper and
lower a.s. instabilities meaningful definitions,
and when the filtering system is one--step observable, we showed that the absolutely upper and lower a.s. stabilities can also be interpreted using  a zero--one law.
To answer the ``zero or one'' question, we presented stability conditions for {general} LTI systems. When the system is one--step observable, it was further shown that
absolutely a.s. stability is equivalent to a.s. one, both of which are
guaranteed by a necessary and sufficient condition in terms of packet arrival rates. Finally, for the so--called non--degenerate systems, a necessary and sufficient upper a.s. stability condition was given.

\end{document}